\documentclass[a4paper]{elsart}

\usepackage{amsmath}
\usepackage{natbib}
\usepackage{hyperref}

\usepackage{graphicx}
\usepackage[english]{babel}
\usepackage{amsmath,amsfonts,latexsym,amssymb}

\graphicspath{{figures/}}

\begin{document}

\begin{frontmatter}

\title{Dynamic Behaviour of Connectionist Speech Recognition with
  Strong Latency Constraints}
\thanks[titlethanks]{This research was carried out at the Centre for Speech
    Technology supported by Vinnova (The Swedish Agency for Innovation
    Systems), KTH and participating Swedish companies and organisations.
}
\author{Giampiero Salvi}
\address{KTH (Royal Institute of Technology),\\
  Dept. of Speech, Music and Hearing,\\        
  Lindstedtsv. 24, 10044 Stockholm, Sweden\\
  \tt giampi@kth.se}
\begin{abstract}
This paper\footnote{Please cite as:\\Giampiero Salvi, ``Dynamic Behaviour of Connectionist Speech Recognition with Strong Latency Constraints'', Speech Communication Volume 48, Issue 7, July 2006, Pages 802-818. \url{https://doi.org/10.1016/j.specom.2005.05.005}} describes the use of connectionist techniques in phonetic
speech recognition with strong latency constraints. The constraints
are imposed by the task of deriving the lip movements of a synthetic
face in real time from the speech signal, by feeding the phonetic
string into an articulatory synthesiser. Particular attention has been
paid to analysing the interaction between the time evolution model
learnt by the multi-layer perceptrons and the transition model imposed
by the Viterbi decoder, in different latency conditions. Two
experiments were conducted in which the time dependencies in the
language model (LM) were controlled by a parameter. The results show a
strong interaction between the three factors involved, namely the
neural network topology, the length of time dependencies in the LM and
the decoder latency.

\end{abstract}


\end{frontmatter}

\section{ Introduction }
This paper describes the use of a hybrid of artificial neural networks/hidden Markov models (ANNs/HMMs) in a speech recognition
system with strong latency constraints. The need for such a system
arises from the task of classifying speech into a sequence of phonetic/visemic
units that can be fed into a rule system to generate synchronised lip
movements in a synthetic talking face or avatar
\citep{gs:Beskow2004}. As the aim is to enhance telephone
conversation for hearing-impaired people \citep{gs:KarlssonEtAl2003},
the total latency allowed between incoming speech and facial animation
is limited by the turn taking mechanism to less than 200 ms. This
includes the latency in capturing the sound and generating and
animating the facial movements. The constraints imposed on the
recogniser are thus especially demanding if compared to other applications of
speech recognition.

In such conditions, and more in general in real-time applications,
conventional decoders based on different flavors of the
Viterbi algorithm \citep{gs:Viterbi1967}, can only be used in an
approximate fashion. This is because the need for incremental results
requires the best-path solution to be based on partial decisions,
with limited look-ahead in time. The difference between the standard
Viterbi solution and the approximated solution is often called
``truncation error''. Truncation error in
the Viterbi algorithm has been extensively studied for convolutional
codes in the area of speech coding
\citep{gs:Kwan1998,gs:Weathers1999}. There, given the relatively
simple nature of the problem, error bounds could be found analytically
and confirmed empirically.

In speech recognition, a few empirical studies dealing with this problem
can be found in the area of broadcast news recognition/transcription
\citep[e.g.][]{gs:Imai2000,gs:Ljolje2000}. In \citet{gs:Ljolje2000} a
system based on incremental hypothesis correction was shown to
asymptotically reach the optimal MAP solution. In
\citet{gs:RobinsonEtAl2002} connectionist techniques are employed in
the same task. These studies are concerned with large vocabulary word
recognition, and have less stringent latency constraints.

The aim of the current study is to analyse the
effect of truncation errors at very low latencies (look-ahead $<$
100 ms) in different set-ups of the language model, while keeping phonetic
recognition as the main focus for the application we have in mind. In
connectionist speech recognition it is of particular interest to study the
interaction between the time evolution model learnt by the time
delayed or recurrent neural networks and the transition model imposed by the
Markov chain, with varying look-ahead lengths.

The first experiment in this study does this by
gradually changing the recognition network from a free loop of phones
(short time dependencies) to a loop of words with increasing
lengths. In the second experiment the language model (LM) is composed of a scaled
mixture of a free loop of phones and a forced alignment topology
where the time dependencies are as long as the utterance. The
gradual modification of the LM is achieved in this second case by changing the mixing
parameter. In addition, a
confidence measure particularly suitable for connectionist models
\citep{gs:WilliamsAndRenals1999} is discussed. The reader also is referred
to \citet{gs:Salvi2003} for an evaluation of the Synface recogniser in more realistic conditions.

The paper is organised as follows: a formal definition of the problem
is given in Section~\ref{Sec:probdef}, Section~\ref{Sec:method}
introduces the method, Section~\ref{Sec:data} the data and
experimental settings. Results are described in
Section~\ref{Sec:results} and discussed in
Section~\ref{Sec:discussion}. Section~\ref{Sec:conclusions} concludes
the paper.
\section{Problem definition and notation}
\label{Sec:probdef}
\subsection{Speech production in formulae}
\begin{figure}
\centering
\includegraphics[scale=0.5]{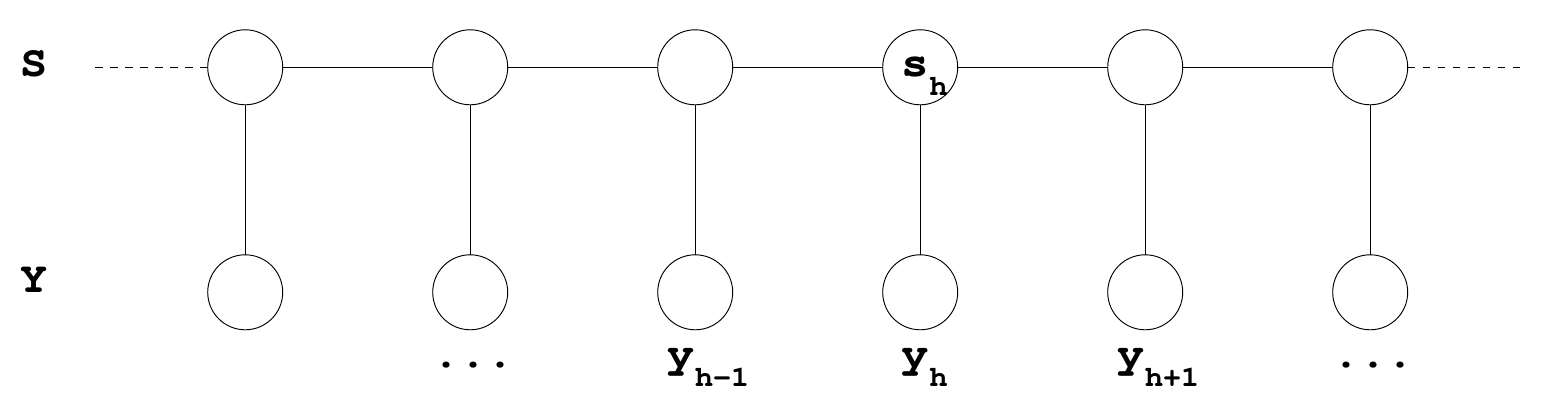}
\caption{Dependencies in a first order HMM represented as a
  \emph{Bayesian network} graph} \label{Fig:HMMDep}
\end{figure}
The process of speech production could be seen as one of encoding
a sequence of symbols $X_1^M=(x_1,\cdots,x_M)$ into a sequence of states
$S_1^N=(s_1,\cdots,s_N)$ with
an associated output sequence $U_1^T=(u_1,\cdots,u_T)$. In our
oversimplified description, $X_1^M$ could represent phonetic classes,
$S_1^N$ are motivated by the dynamics introduced by articulatory
gestures that in turn generate the speech signal $U_1^T$. Phonetic
speech recognition is therefore the process of recovering the original
sequence $X_1^M$ on the base of some features $Y_1^N=(y_1,\cdots,y_N)$
extracted from $U_1^T$. 
When the feature extraction procedure is assumed to be given, as in the
current study, the distinction between $U$ and $Y$ is not
essential. Speech production is then a (stochastic) function of
the kind: $P:X \to Y$. The natural choice for characterising this
function is a Markov model $\Theta$ where the states $s_i$ are assumed to vary
synchronously with the features $y_j$, which explains why we indicated
the length of $S$ and $Y$ with the same symbol $N$.
Besides an \textit{a~priori} term, $\Theta$ is then fully specified by
the distribution of state transition probabilities $a_{ij} =
P(s_j|s_i)$ and the likelihood of the data generation given a certain
state $b_i(Y_h^k) = P(Y_h^k|s_i)$.
The usual assumption is to consider the latter as local models, in the
sense that the state $s_i$ at a particular time $h$ influences the
observation only at that time $h$: $P(Y_h^k|s_i) = P(y_h|s_i)$, as
illustrated in Figure~\ref{Fig:HMMDep}.
In this case, all the information about the dynamic evolution of the
process under study is coded in the transition model $a_{ij}$.
\subsection{State-to-output probability estimators}
\begin{figure}
\centering
\includegraphics[scale=0.5]{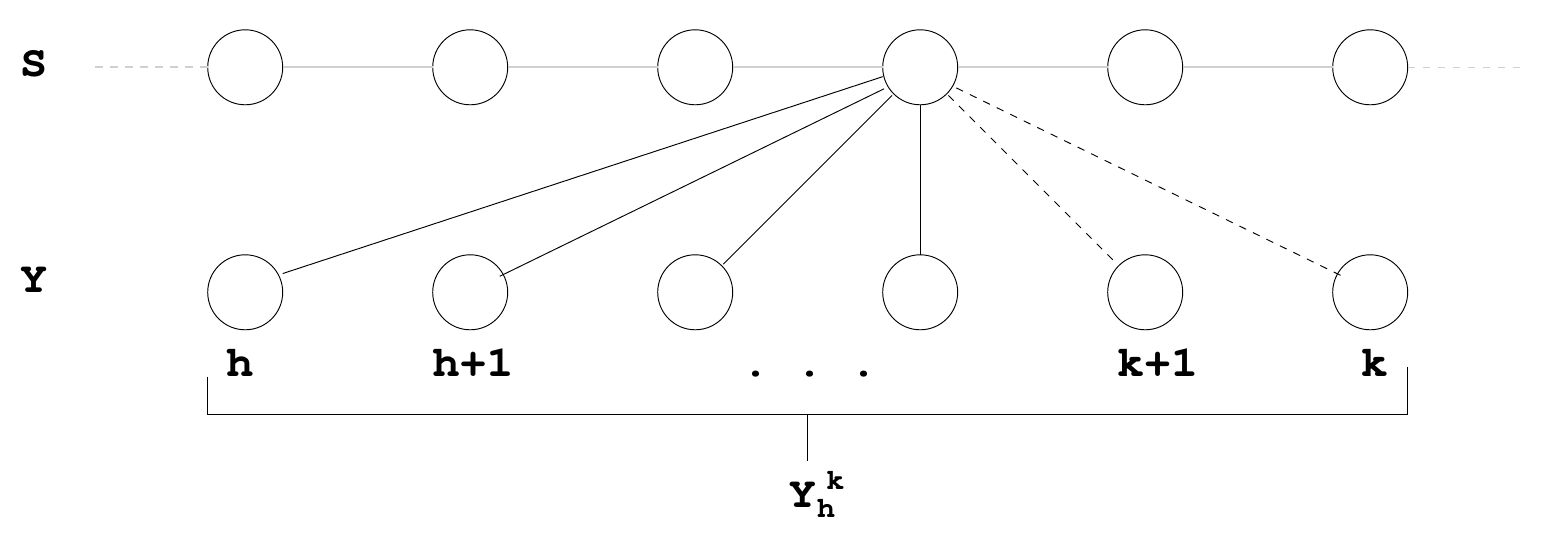}
\caption{Dependencies introduced by time dependent MLPs.} \label{Fig:RNNDep}
\end{figure}
\label{StateToOut}
\citet{gs:Robinson1994} has shown how multi layer
perceptrons (MLPs) can be efficient estimators for the
\textit{a~posteriori} probabilities $P(x_i|Y_1^n)$ of a certain state
$x_i$ given an observation $Y_1^n$. A particularly
efficient training scheme uses back propagation through time
\citep{gs:Werbos1990} with a cross entropy error measure
\citep{gs:Bourlard1993}. If the nonlinearity in the units is in the
$\tanh$ form, we can write for the state to output probabilities:
\begin{equation}
P(Y_h^k|x_j) = \frac{P(x_j|Y_h^k)P(Y_h^k)}{P(x_j)} \simeq
\frac{o_j+1}{2}\ \frac{P(Y_h^k)}{P(x_j)}
\label{Eq:estimate}
\end{equation}
Where $x_j$ is a phonetic
class and $o_j$ the activity at the output node corresponding to that
class. The linear transformation in the formula ($(o_j+1)/2$) is
necessary to transform the $\tanh$ values, that span from -1 to 1,
into probabilities. In the following we will refer to output
activities of the MLP as the linearly transformed outputs that assume
values in the range $[0,1]$. $Y_h^k$ is the sequence of feature vectors spanning a
window of time steps that depends on the dynamic properties of the
MLP. In the case of simple feed-forward nets, $Y_h^k$ reduces to the
current frame vector $y_k$, while for strict recurrent topologies (RNN),
$h=1$ and $k$ is the current frame. This is illustrated in
Figure~\ref{Fig:RNNDep} that shows how the dynamic properties of the
neural network can introduce dependencies between states and
observations that span over a number of time steps. In
\citet{gs:Stroem1992} a mixture of time delayed and recurrent
connections was proposed. In this model the input layer received
contributions both from the past and the future frames thanks to time
delayed connections with possibly negative delays (represented in the
Figure by dashed lines).
In this study, only positively delayed connections are considered, in
order to reduce the total latency of the system.
\subsection{Interaction between HMM topology and ANN dynamic
  properties} \label{Sec:Dynamics}
\begin{figure}
\centering
\includegraphics[scale=0.5]{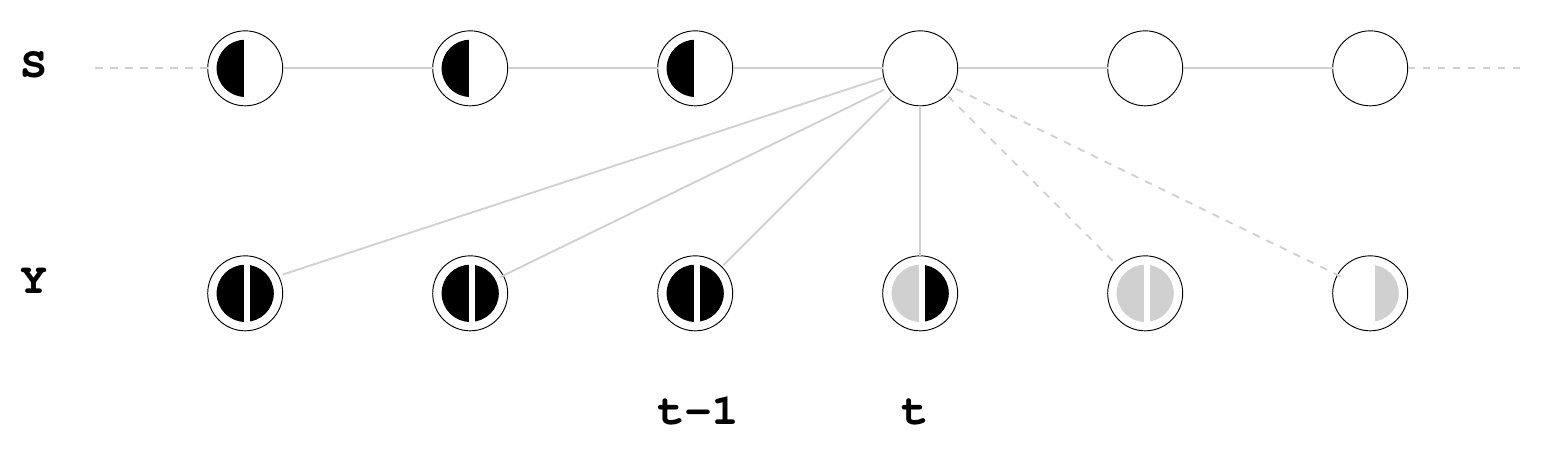}
\caption{Interaction between the $\delta$ and $b$ terms in Viterbi decoding.} \label{Fig:HMMANNInter}
\end{figure}
Given the probabilistic model $\Theta$, the
\textit{maximum~a~posteriori} (MAP) solution to the speech recognition
problem is the sequence $X_1^M$ that maximises 
\begin{equation}
P(X_1^M|Y_1^N,\Theta)=P(x_1,\cdots,x_M|y_1,\cdots,y_N,\Theta)
\label{Eq:mapsolution}
\end{equation}
A more pragmatic solution, provided by the Viterbi
algorithm, approximates the sum over all possible state sequences,
implicit in Equation~\ref{Eq:mapsolution}, with a maximum operation. Since in our
model $X_1^M$
is fully determined by $S_1^N$, the recognition problem is equivalent
to finding the sequence $S_1^N$ for which $P(S_1^N|Y_1^N,\Theta)$ is maximum.
This can be done iteratively according to the Viterbi recursion formula:
\begin{displaymath}
 \delta_t(j) = \max_{i}[\delta_{t-1}(i)a_{ij}]b_j(y_t)
\end{displaymath}
Where $b_j(y_t) = P(y_t|x_j)$ is the likelihood of the current observation
given the state and $\delta_t(j)$ is the Viterbi accumulator. In
practice we substitute to $b_j(y_t)$ the
estimate of $P(Y_h^k|x_j)$ given by Equation~\ref{Eq:estimate}. In the
case of recurrent MLPs, $P(Y_h^k|x_j)$ equals $ P(Y_1^t|x_j)$ and the
information contained by $\delta_{t-1}(i)$ and $b_j(Y_1^t)$ in the
Viterbi recursion formula becomes
widely overlapping. Figure~\ref{Fig:HMMANNInter} illustrates
this by indicating which states and observations the two terms in the
Viterbi recursion formula depend on. Left semicircles refer to the
term $\delta_{t-1}(i)$ and right semicircles to the term
$b_j(Y_1^t)$. Grey semicircles are included if the MLP has negatively
delayed connections.

As the two sources of information in the Viterbi recursion are
strongly dependent, we expect the evidence brought by their joint
contribution to be lower than the sum of each single contribution,
as if they were independent.

\subsection{Interaction between HMM topology and look-ahead length }
\label{Sec:Truncation}
When truncation is considered, the optimal solution at time step $n$ is the
state $s_n$ extracted from the sequence
$S_1^{n+D}=(s_1,\cdots,s_n,\cdots,s_{n+D})$ that maximises
$P(S_1^{n+D}|Y_1^{n+D},\Theta)$, where $D$ denotes the look-ahead
length in time steps.
\begin{figure*}
  \centering
  \includegraphics[scale = 0.5]{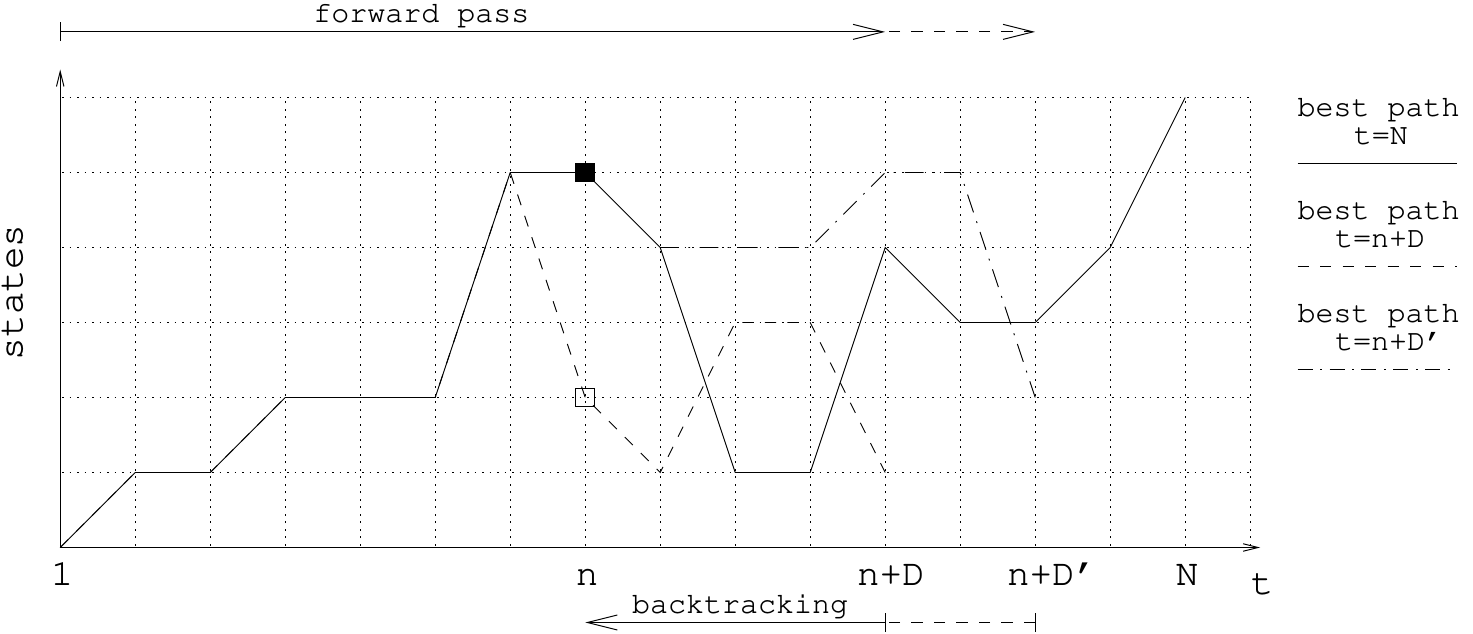}
  \caption{Trellis plot in with three Viterbi paths (varying
  look-ahead length)}
  \label{Fig:Search}
\end{figure*}
The difference between the two approaches is
exemplified in Figure~\ref{Fig:Search}. The grid displays the states
as a function of time (trellis). The continuous line shows the Viterbi
solution, while the dashed and dashed-dotted lines refer to the best
path obtained using the partial information up to $t=n+D$ and
$t=n+D^\prime$, respectively. The figure also illustrates a
phenomenon that is common in practice: the influence of an observation
at time $t_1$ over the result at time $t_2$ decays with the distance
$D = |t_1-t_2|$. In the example the observations in the interval
$[n\!+\!D\!+\!1,n\!+\!D^\prime]$ influence the result at time $n$, as
prolonging the look-ahead from $D$ to $D^\prime$ leads to different results
(open and filled squares). With respect to the solution at time $n$,
however, adding the observations in $[n\!+\!D^\prime\!+\!1,N]$ to the
search mechanism does not change the response. As a result the
truncated solution will in general asymptotically approach the
standard Viterbi solution (filled square in this case) as $D$ increases.
The value $D^*$ at which the two solutions become
indistinguishable depends on the dynamic characteristics of the problem at
hand, i.e. on the time correlations in $Y$ and
on those imposed by the transition model $\Theta$.


\section{Method} \label{Sec:method}
To evaluate the interaction between the language model, the properties
of the probability estimators, and truncation in the Viterbi decoder,
three-factor experiments were designed. The factors involved are: the
length of time dependencies in the recognition network
(\emph{language model}), the dynamical properties of the \emph{probability
  estimators} and the \emph{look-ahead length}.

\subsection{Language model}
Varying the length of time dependencies in the language model (LM)
was simulated in two different ways. In both cases, a different
LM is used for each utterance, based on information
contained in its transcription.

The first method consists of constructing the recognition network as
the union of two topologies with transition weights scaled by
a factor $\alpha$:
\begin{equation}
 \mbox{LM}(\alpha) = \alpha \ \mbox{AL} \ \cup \ (1-\alpha) \ \mbox{PL}
\end{equation}
PL specifies a free loop of the phones included
in the transcription for each utterance, while AL is the
LM obtained by connecting in a sequence the phones specified by
the transcription of each utterance. When $ \alpha \to 0 $ the grammar
has the short time dependencies of a phone loop, when $ \alpha \to 1$
the time dependencies are as long as the utterance, and the
recogniser performs forced alignment. The parameter $\alpha$
assumes seven values in the experiments:
$0,0.1,0.3,0.5,0.7,0.9,1$. This design will be referred to as
the ``alpha test''.

\begin{figure}
  \centering
  \includegraphics[scale = 0.5]{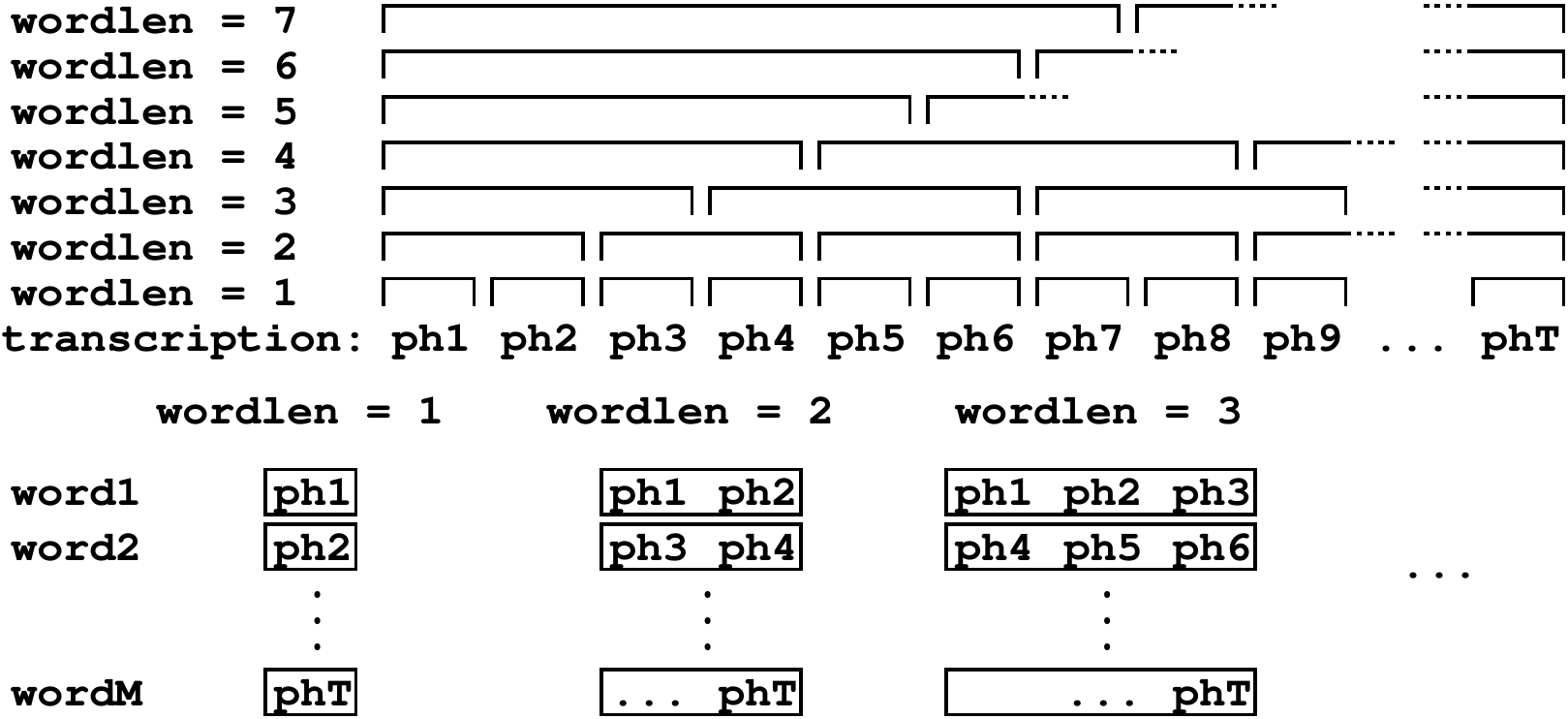}
  \caption{Illustration of the ``wordlen test'' design: the transcription of
  each test utterance is spit into words of increasing lengths, that are
  used in the recognition network.}
  \label{Fig:WordLenTest}
\end{figure}
In the second method, the LM defines a loop of words, where a word is
defined by successively extracting $N$ phones from the transcription of
each utterance (see Figure~\ref{Fig:WordLenTest}). For $N = 1$ the LM is again a loop of phones. The
parameter $N$ assumes the values from 1 to 7. To be
noted here is that, since each phone is modelled by a three state
Markov model, the lower bound to the time dependencies induced by this
LM ranges from 3 to 21 frames in the experiments. This design
will be referred to as the ``wordlen test''.

As already noted, the phone-loop condition ($\alpha = 0$ or $N=1$) is
obtained in the alpha and wordlen tests by selecting for each test utterance
only those phones contained in its transcription. This was necessary to
guarantee homogeneity with the other conditions ($\alpha \neq 0$ or $N
\neq 1$), as the main objective of the experiment is to compare the
different conditions, rather than computing an absolute performance score.
When evaluating the recogniser from the
perspective of the application, a loop of all phones should be used. This
condition is analysed in Section~\ref{Sec:results} as a reference.

A factor that turned out to be important in the evaluation of the
different LMs, especially in the low-latency condition, is the
phone-to-phone transition probability. In the phone-loop condition the
transition probability from phone $i$ to phone $j$ is $1/N$ (the uniform
case) where $N$ is the number of phones. In the other
conditions, the within-word phone-to-phone transition probability
should in principle be 1, as each phone can only be followed by the
next in the pronunciation model. This introduces terms that turned out
to strongly
affect the search mechanism, especially for low-latency
conditions. The solution was to keep the same transition probabilities
in both cases ($1/N$), releasing the constraint of a stochastic
grammar (outgoing transition probabilities that sum to 1).
This is a common practice in speech recognition where multiplicative
and additive terms are usually applied to a subset of the transition
probabilities, often corresponding to the language model. The aim is
however different, as we are not directly interested in tuning the
performance of the decoder, but rather in ruling out from the comparison
factors that do not depend on the dynamic properties of the LM.
\subsection{Probability estimators}
The second experimental factor is the ability of
the state-to-output probability models to express time variations. In
this case, similarly to \cite{gs:Salvi2003}, three multi layer
perceptrons (MLPs) were used with different complexities and dynamic
properties. One feed-forward MLP is considered as a static model,
while two recurrent MLPs represent models capable of learning the time
evolution of the problem (details in Section~\ref{Sec:acousticmodels}).

\subsection{Look-ahead length}
\label{Sec:lookahead}
The third experimental factor is the look-ahead length $L$
that can be varied in our decoder. One problem is how to decode the
last $L$ frames at the end of each utterance. As every utterance
begins and ends with silence, it was suggested in \citet{gs:Salvi2003}
to decode the whole test set in a continuous stream, limiting the
boundary problem only to the last file. Here, in contrast, each
utterance is analysed separately and the result for the last $L$
frames is assumed equal to the best path obtained when the forward
phase has reached the end of the utterance. This is a somewhat more
standard way of computing the Viterbi solution and was necessary to
enable the use of a different language model for each utterance. Five values
of the look-ahead length (in frames) were used in the experiments:
1,~3,~5,~10,~20.

\subsection{Scoring method} 
The scoring method chosen for this study is frame-by-frame correct
classification rate simply computed as the ratio between the number of
correctly classified frames and the total number of frames. A correct
classification occurs when a frame has been assigned the same phonetic
class as in the transcription. This method was
preferred to the more common minimum edit distance at the symbolic
level, because the application we have in mind requires not only
correct classification of the speech sounds, but also correct segment
alignment. In Section~\ref{Sec:results}, phone-loop results are
reported in terms of frame-by-frame correct classification rate, as
well as accuracy and percent of correct symbols \citep{gs:Young2002}
in order to compare these scoring methods.

\subsection{Confidence measure}
\label{Sec:confidence}
As noted in \citet{gs:WilliamsAndRenals1999} the property of
multi-layer perceptrons to estimate posterior probabilities, as
opposed to likelihoods, is advantageous when computing confidence
measures of the acoustic models. A simple measure of the acoustic
confidence is the per-frame entropy of the $k$ phone
class posterior probabilities. Although the
entropy measure, in \citet{gs:WilliamsAndRenals1999}, is averaged over
a number of frames, we will consider a frame-by-frame measure.
A factor that strongly influences the entropy measure is the choice
of the target values during training of the networks. A common practice
is to use $0+\epsilon$ and $1-\epsilon$, with small $\epsilon$, as
target values, in order to
speed up the convergence in the standard back propagation
algorithm (Note that when using $\tanh$ squashing functions, the
limits are internally -1 and 1, and the above discussion refers
to the linearly transformed outputs of the network, see also
Equation~\ref{Eq:estimate}). As a consequence, the networks trained this
way are more noisy in the sense that the activities of the inactive
output nodes seldom fall below $0+\epsilon$. Strictly speaking, this
also gives incorrect posterior probabilities estimates.

To show the effect of the target values on the entropy measure we
consider a simplified example. We assume that the activities of the
output nodes of the network trained with target values $o_H$ (active)
and $o_L$ (inactive), can only assume those values when the network is
excited by new input. In reality the activities take any value in between and
sometimes even outside the range $[o_H,o_L]$. Even if the network is
trained with only one active output node per time step, there will be,
during excitation, a number $N_H$ of simultaneously active nodes. If we
call $N_L$ the number of inactive nodes ($N_H+N_L=N$ is the total
number of output nodes), then, from the definition of the entropy:
\begin{figure}
  \centering
  \includegraphics[scale = 0.6]{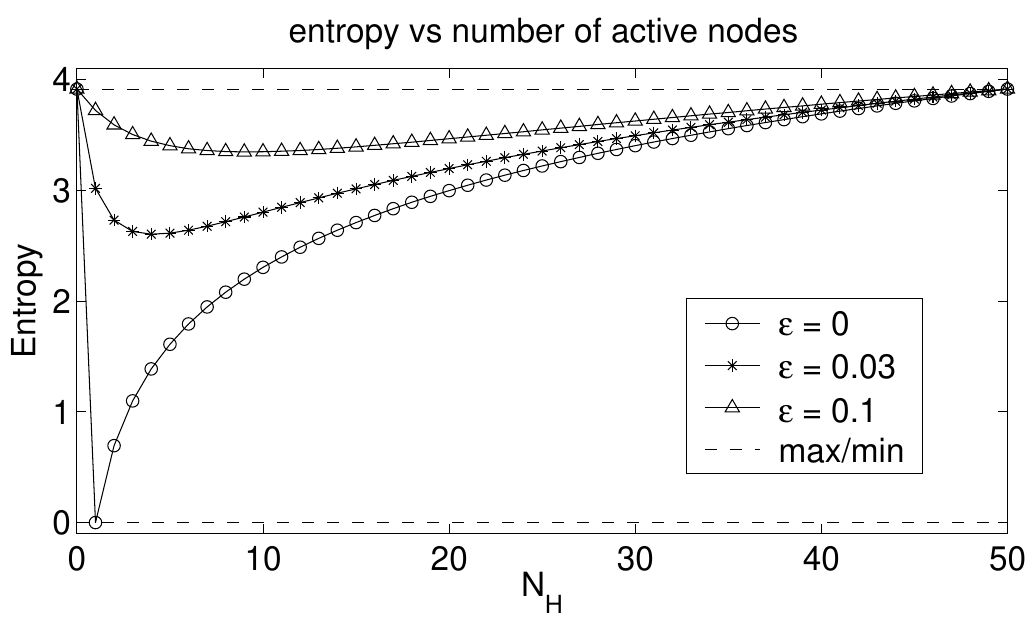}
  \caption{Simulation of entropy of a distribution with $N_H$ high levels}
  \label{Fig:EntropySim}
\end{figure}
\begin{equation}
 H= -N_H b_H \log b_H -N_L b_L \log b_L
\label{Eq:entropysimulation}
\end{equation}
where $b_H$ and $b_L$ are the normalised activities obtained imposing
that the sum of probabilities be 1:
\begin{displaymath}
b_H = \frac{o_H}{N_H o_H + N_L o_L},\ \ b_L = \frac{o_L}{N_H o_H + N_L o_L}
\end{displaymath}
In the case of symmetrical values, i.e. $(o_L,o_H) =
(\epsilon,1-\epsilon)$, Equation~\ref{Eq:entropysimulation}
 assumes the form:
\begin{displaymath}
 H = \log (N_H (1-\epsilon) +N_L \epsilon) - 
\frac{N_H(1-\epsilon) \log (1-\epsilon) + N_L \epsilon \log \epsilon}
{N_H (1-\epsilon) + N_L \epsilon}
\end{displaymath}
When $\epsilon \to 0$ ($o_H \to 1$ and $o_L \to 0$), the entropy $H$
tends to $\log(N_H) $, as easily seen in the formula.
In Figure~\ref{Fig:EntropySim} the entropy is plotted as a function of
the number of active nodes $N_H$, for the cases $\epsilon = 0, 0.03,
0.1 $, and for $N=N_H+N_L=50$ as in our networks.
The figure shows how the entropy of the output of a network trained
between 0 and 1, given our assumptions, spans the whole range from 0
to $\log N$, while the network trained between
$\epsilon$ and $1-\epsilon$ has a more limited range in the high
entropy region. The range strongly depends on $\epsilon$.
In our experiments one network was trained with $[0,1]$ targets and the
other two with $[0.1,0.9]$ targets ($\epsilon=0.1$).

In Section~\ref{Sec:ConMeasResults} the impact of $\epsilon$ on
the entropy based confidence measure is discussed with examples on the
test data.

\section{Data} \label{Sec:data}
The experiments were performed on the Swedish SpeechDat corpus
\citep{gs:Elenius2000}, containing recordings of 5000 speakers over
the telephone.
The official training and test sets defined in SpeechDat and
containing respectively 4500 and 500 speakers, were used in the
experiments. Mel frequency cepstrum features were extracted at 10 ms
spaced frames.

\begin{figure}
  \centering
  \includegraphics[scale = 0.8]{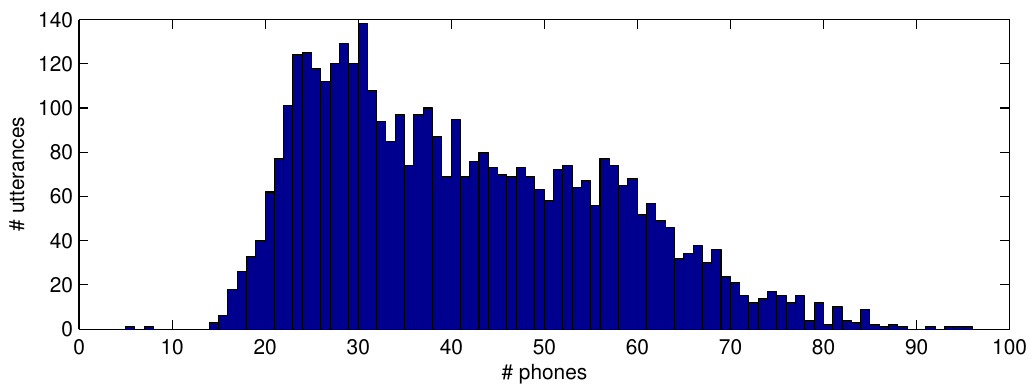}
  \caption{Distribution of the length of the test utterances in terms
  of number of phonemes.}
  \label{Fig:TestTransLen}
\end{figure}

The training material for the neural networks, and the test
material were restricted to the phonetically rich sentences. The
training material was further divided into training and
validation sets of 33062 and 500 utterances respectively. The test set
contains 4150 utterances with an average length of 40.9 phonemes per
utterance. Figure~\ref{Fig:TestTransLen} shows the distribution of the
length of the test utterances in terms of number of phonemes.

One problem
with the SpeechDat database, that is important when training the MLPs
and for testing at the frame level, is the unbalanced amount of
silence frames compared to the amount of material for any
phonetic class. Part of the silence frames that were concentrated at the
beginning and at the end of each utterance, was removed.

Since the dataset lacks phonetic transcriptions, some
pre-processing was necessary. The time-aligned reference, used
for both training and testing the MLP models, was obtained
with forced alignment employing the word level transcriptions, the official
SpeechDat lexicon and triphonic HMMs based on Gaussian models. The
HMMs were trained with the procedure defined in the \mbox{RefRec} scripts
\citep{gs:Lindberg2000}. The alignment lattice allowed non speech
sounds to be inserted between words. The method proved to be
reasonably accurate for our purposes.

\subsection{Acoustic Models} 
\label{Sec:acousticmodels}
Three multi-layer perceptrons were used in this study. The first
({\tt ANN}) is a feed-forward network with two hidden layers of 400
fully connected units each. {\tt RNN1} and {\tt RNN2} are recurrent
networks with one hidden layer of 400 units and a varying number of time
delayed connections. The choice of topology in {\tt ANN} aimed at
ensuring a comparable complexity (in number of free parameters) between
{\tt ANN} and {\tt RNN1}.

As a reference, results obtained with a set of Gaussian mixture
models ({\tt GMM}), are reported together with the multi-layer
perceptrons results. The {\tt GMM} results show the discriminative
power of the set of Gaussian models, when the phonetic class with the
maximum \emph{a posteriori} probability (MAP) is selected for each
frame. The Gaussian mixture parameters where estimated using the HMM
training with 32 Gaussian terms per state.

\begin{table}
\caption{Details on the Acoustic Models and Frame-by-frame MAP results} \label{Tab:gmmann}
\begin{tabular}{lccccc} \hline \hline
 model & \# params & \# hidd.units & \# hidd.layers & recurrent &
 \% correct frame \\ \hline
 {\tt GMM} & 379050 & - & - & - & 35.4\% \\
 {\tt ANN} & 186050 & 800 & 2 & no & 31.5\% \\
 {\tt RNN1} & 185650 & 400 & 1 & yes & 49.6\% \\
 {\tt RNN2} & 541250 & 400 & 1 & yes & 54.2\% \\ \hline \hline
\end{tabular}
\end{table}
Details on the acoustic models are reported in
Table~\ref{Tab:gmmann}. The table shows the overall number of parameters, and,
in the case of the perceptrons, the number of hidden layers and hidden
units and the dynamic characteristics. The last column reports the
correct frame classification rate when the maximum \emph{a posteriori} class
was selected frame-by-frame.
\subsection{Implementation note}
The HMM training was performed using the
HTK Toolkit \citep{gs:Young2002}. The MLP training algorithm was
implemented in the NICO Toolkit~\citep{gs:StroemWEB}. The modified Viterbi
algorithm is the decoder used in the SYNFACE project
\citep{gs:Salvi2003,gs:KarlssonEtAl2003}, and, together with the other
tools used in the experiments, was implemented by the
author. The statistical analysis was performed using the R software
\citep{gs:R-project}. All experiments were performed on a GNU-Linux
platform running on standard hardware (PC).

\section{Results} \label{Sec:results}
\begin{figure}
  \centering
  \includegraphics[scale = 0.42]{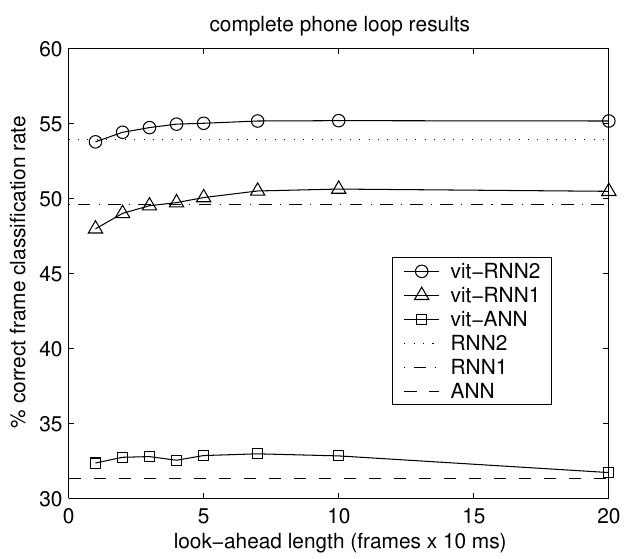}
  \includegraphics[scale = 0.42]{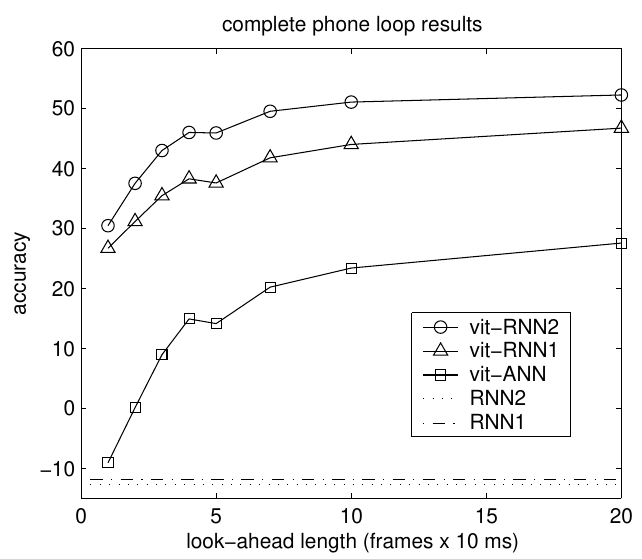}
  \includegraphics[scale = 0.42]{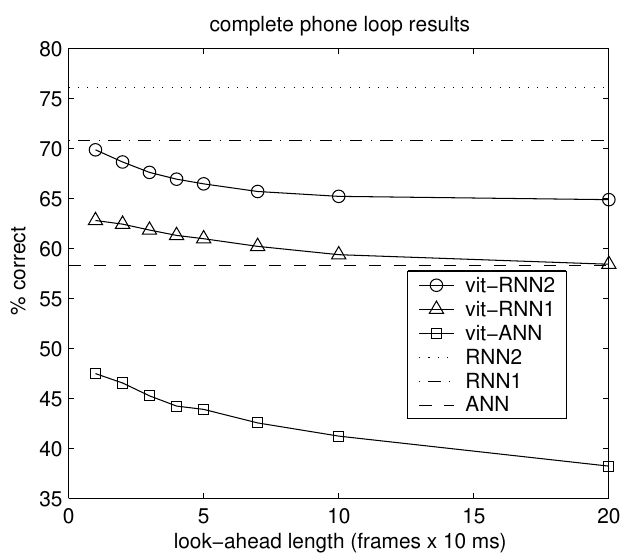}
  \caption{Phone loop results}
  \label{Fig:phoneloopresults}
\end{figure}
\begin{figure}
\centering
\includegraphics[scale=0.8]{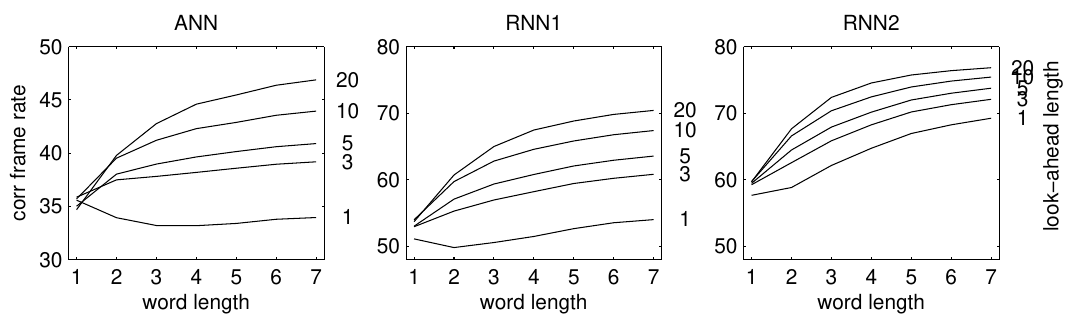}
\includegraphics[scale=0.8]{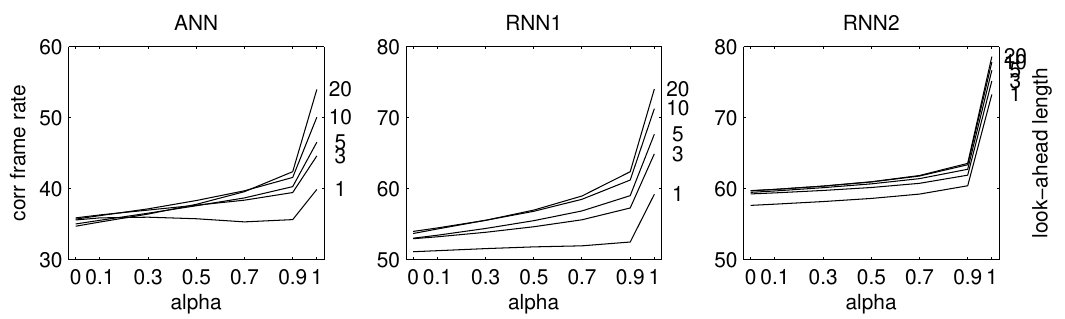}
\caption{Summary of ``wordlen'' (top) and ``alpha'' (bottom) results} \label{Fig:resultSummary}
\end{figure}
Results obtained with a normal phone loop
are reported in Figure~\ref{Fig:phoneloopresults} as a reference
to the performance of the recogniser in the real task. The left plot
in the figure shows the average correct frame classification rate over
the 4150 test utterances for varying look-ahead length and for the three neural
networks ({\tt vit-ANN}, {\tt vit-RNN1} and {\tt vit-RNN2}). The
horizontal lines in the figure indicate the classification rate
without Viterbi decoding, i.e. selecting the highest activity output at
each frame (frame-by-frame maximum a posteriori). The results are very
close to the ones obtained in \citet{gs:Salvi2003}, the differences
being due to the way the boundary effects are handled (see
Section~\ref{Sec:lookahead}), and to the fact that in
\citet{gs:Salvi2003} a global score was computed over the whole test
material, while here we compute the correct frame classification rate
of each utterance and then average the results.

The middle and right plots in Figure~\ref{Fig:phoneloopresults} show
the accuracy and percent of correct words as defined in
\citep{gs:Young2002}. These results are reported in order to compare
the three scoring methods, and to mention that none of them are fully
satisfying given the application. Accuracy and correct words do not
take into account segment boundary alignment in time and were
therefore discarded in the following evaluation. Correct frame
classification rate, in contrast, does not indicate how stable the
result is (number of symbol insertions).

The ``wordlen'' and ``alpha'' tests results are summarised in
Figure~\ref{Fig:resultSummary}.
In the first case (top) the average correct frame
rate is plotted as a function of the word length in the ``wordlen''
test for different
values of the look-ahead length and for the three multi-layer
perceptrons. In the second case (bottom) the $x$-axis indicates the
$\alpha$ parameter in the ``alpha'' test. Note that the range of the
$y$-axis in the {\tt ANN} case is different from the other two.

Longer time dependencies in the language model (LM) and longer look-ahead
lengths are beneficial in most conditions, as most of the curves increase
monotonically and do not overlap. Exceptions to this are
the conditions in which the static model {\tt ANN} is used in
conjunction with either a short time LM or a short look-ahead
length. In those cases, more irregular results can be found (see left
plots in the Figure). Examples are the fluctuating results
corresponding to different look-ahead lengths when wordlen = 1 or
$\alpha=0$ (top-left and bottom-left plots) and the non-monotonic
variation of the score with respect to the word length and $\alpha$
when the look-ahead length is one frame (top-left and bottom-left
plots). The last phenomenon can be found also in the {\tt RNN1} case
(top-middle plot).

In the following, these results will be analysed in details with
statistical means.
\subsection{Wordlen test: short to long time dependencies}
\begin{figure}
\centering
\includegraphics[scale=0.8]{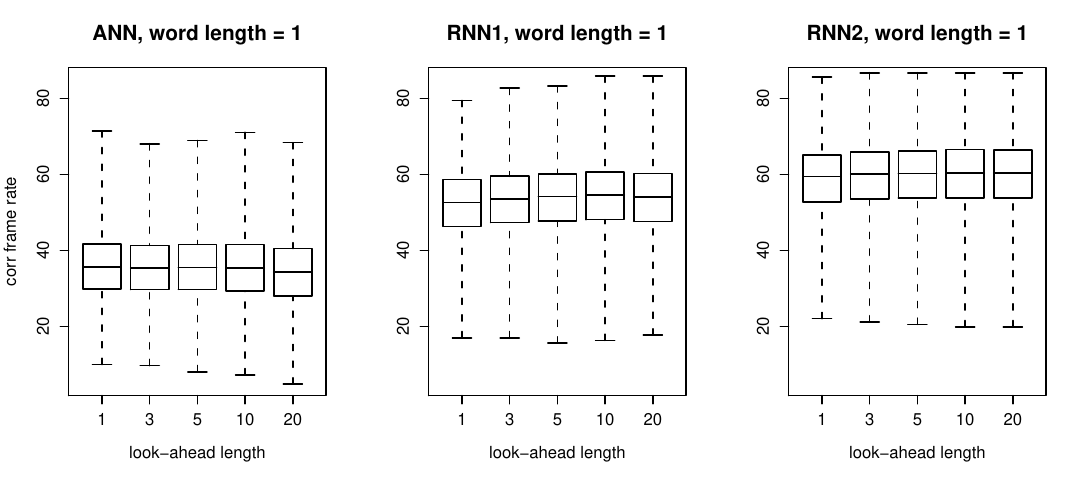}
\includegraphics[scale=0.8]{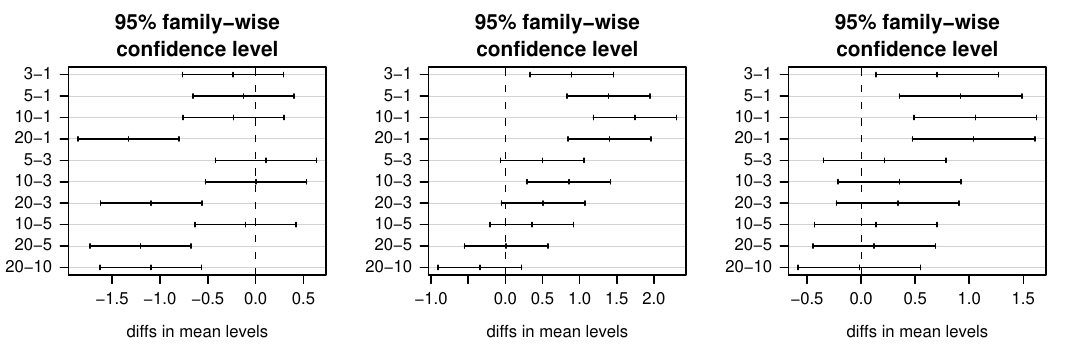}
\caption{Box-plots (top) and 95\% family-wise Tukey confidence intervals (bottom), word length = 1} \label{Fig:wordlen1}
\end{figure}
Figure~\ref{Fig:wordlen1} (top) shows 
box plots for the phone-loop case (word length = 1) for different
look-ahead lengths. The middle line in the box displays the median,
while the lower and higher lines, called lower and higher hinges, display
respectively the first and third quartiles. The lower and higher
whiskers show the full range of the data. It is not clear in the {\tt
  ANN} case, whether the use of longer look-ahead is beneficial to the
recognition score. In {\tt RNN1} and {\tt RNN2} there is a slight
improvement along increasing look-ahead lengths. Applying an ANOVA to
the data returns significant differences in all the three cases
(respectively for {\tt ANN}, {\tt RNN1} and {\tt RNN2}:
F(4,20745) = 15.31, 21.79, 8.96; p = 1.65x10$^{-12}$, $<$2.2x10$^{-16}$, $<$3.1x10$^{-7}$).
A successive Tukey multiple comparison test is
visualised in Figure~\ref{Fig:wordlen1} (bottom). The figure indicates
the 95\% family-wise intervals for all possible combination of the
values of the look-ahead factor. The difference between condition $x$
and $y$ is indicated by $x-y$ that is the increase in correct frame
rate going from look-ahead length $y$ to $x$. If an interval crosses
the zero line, the differences between the two conditions are not
significant.

There are significant differences in {\tt ANN} but with
negative signs. In {\tt RNN1} the difference $10-3$ and all
differences between $L=1$ and $L\ne 1$ are significant. Finally in
{\tt RNN2} only the $L=1$ condition seems to be distinguishable from
the others.

\begin{figure}
\centering
\includegraphics[scale=0.8]{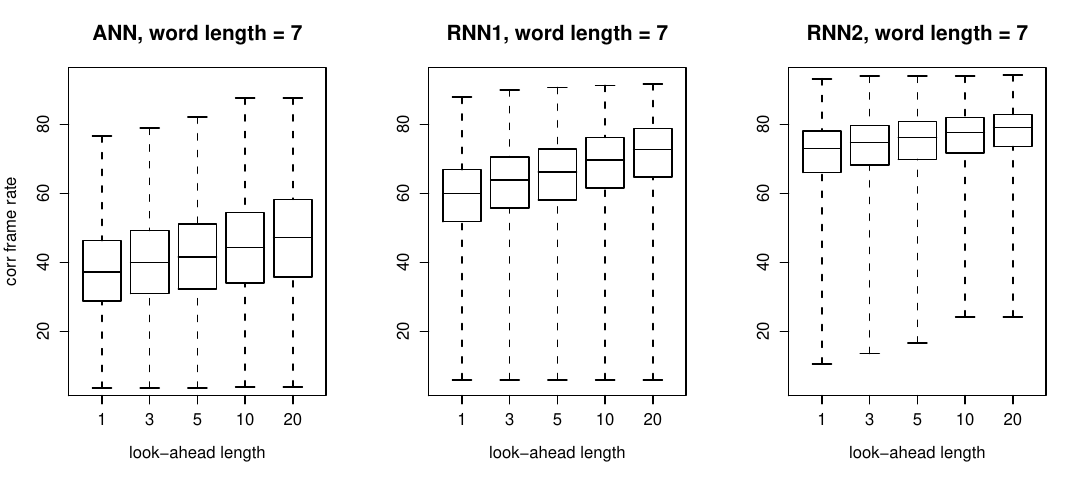}
\includegraphics[scale=0.8]{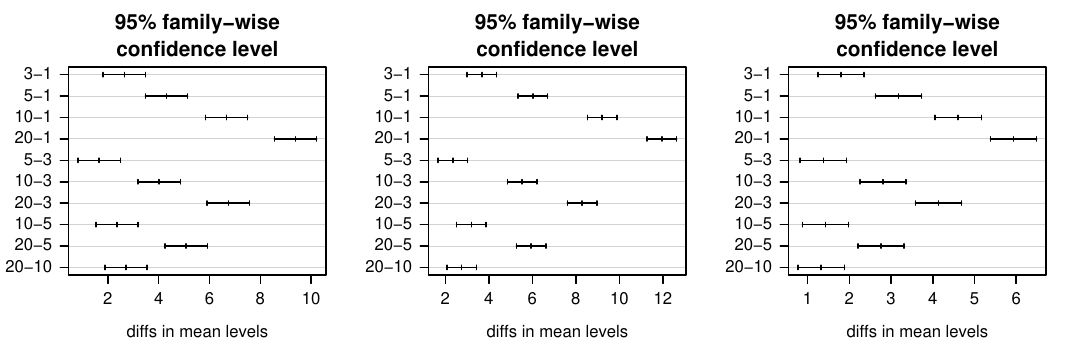}
\caption{Box-plots (top) and 95\% family-wise Tukey confidence intervals (bottom), word length = 7} \label{Fig:wordlen7}
\end{figure}
On the other end of the word length parameter values (wordlen = 7),
the information carried by the transition model, and the
Viterbi processing has a stronger effect on the feed-forward perceptron.
Figure~\ref{Fig:wordlen7} (top) shows the corresponding box plot. The
differences are significant in all cases (respectively for {\tt ANN},
{\tt RNN1} and {\tt RNN2}: F(4,20745) = 281.32, 707.16, 262.17; p =
$<$2.2x10$^{-16}$). Multiple comparison leads to significance in every
difference as illustrated by Figure~\ref{Fig:wordlen7} (bottom).
Of the consecutive distances, $3-1$ is always the greatest. In
{\tt ANN} and {\tt RNN1}, $10-5$ is greater than $5-3$.

Table~\ref{Tab:wordlensummary} summarises the results for
intermediate word lengths. A plus sign in a {\tt x-y} column indicates
a positive significant difference between the latency conditions {\tt
  y} and {\tt x}, a minus sign indicates a significant negative difference and,
finally, ``{\tt n}'' no significant difference.
\begin{table}
\caption{Wordlen test: Tukey HSD multiple comparison results} \label{Tab:wordlensummary}

\begin{scriptsize}
\begin{verbatim}
ANN
wordlen  3-1   5-1   10-1  20-1  5-3   10-3  20-3  10-5  20-5 20-10
1         n     n     n     -     n     n     -     n     -     -
2         +     +     +     +     +     +     +     +     +     n
3         +     +     +     +     +     +     +     +     +     +
4 to 7 equal to 3
RNN1
wordlen  3-1   5-1   10-1  20-1  5-3   10-3  20-3  10-5  20-5 20-10
1         +     +     +     +     n     +     n     n     n     n
2         +     +     +     +     +     +     +     +     +     +
3 to 7 equal to 2
RNN2
wordlen  3-1   5-1   10-1  20-1  5-3   10-3  20-3  10-5  20-5 20-10
1         +     +     +     +     n     n     n     n     n     n
2         +     +     +     +     +     +     +     +     +     +
3 to 7 equal to 2
\end{verbatim}
\end{scriptsize}
\end{table}

\subsection{Alpha test: short to long time dependencies}
\begin{figure}
\centering
\includegraphics[scale=0.8]{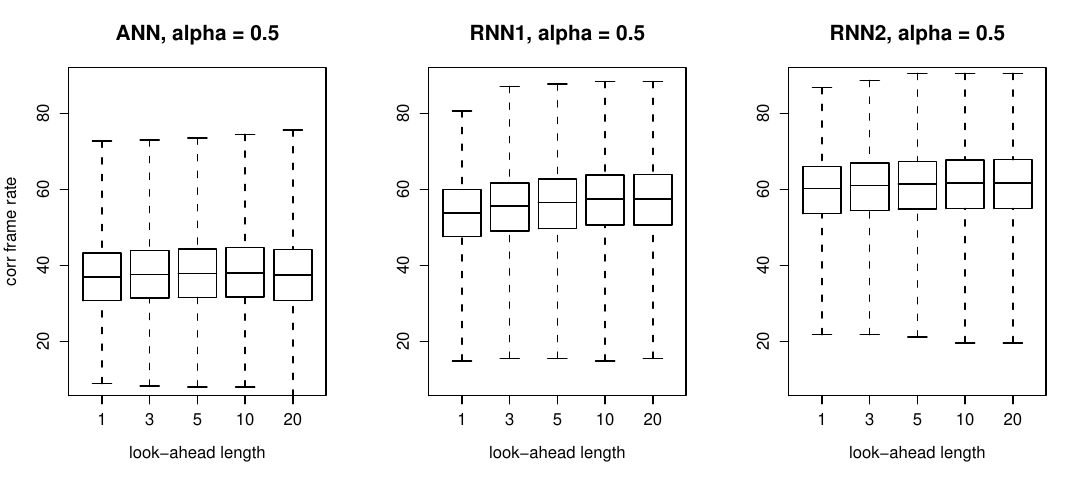}
\includegraphics[scale=0.8]{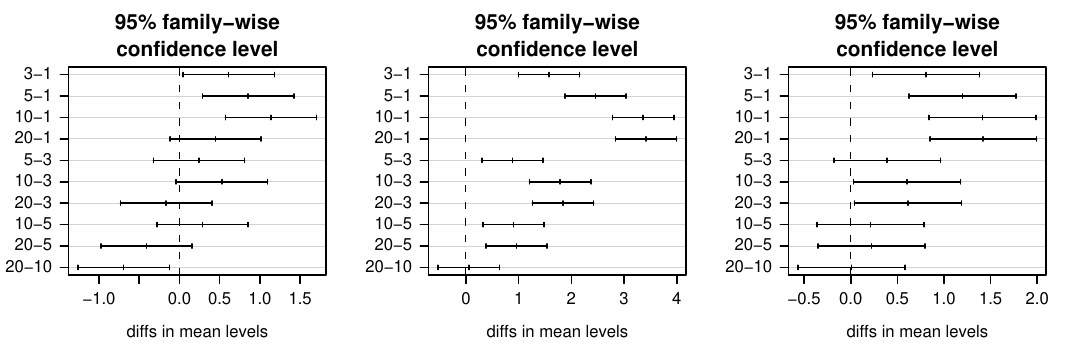}
\caption{Box-plots (top) and 95\% family-wise Tukey confidence intervals (bottom), alpha = 0.5} \label{Fig:alpha0.5}
\end{figure}

Experiments carried out with the ``alpha'' test show similar results
to the ``wordlen'' test. In the phone loop condition ($\alpha=0.0$)
the language model is equivalent to the one in the ``wordlen'' test with word
length 1 (see previous section). Figure~\ref{Fig:alpha0.5} shows
the intermediate condition $\alpha=0.5$. In the {\tt ANN} case
the $3-1$, $5-1$ and $10-1$ differences are significant and positive,
while the $20-10$ difference is significant but negative.
{\tt RNN2} shows clear significant differences when changing
from 1 frame to longer look-ahead. The $5-3$ and $10-3$ differences are also
significant but less evidently. With {\tt RNN1} all differences are
significant except $20-10$.

\begin{figure}
\centering
\includegraphics[scale=0.8]{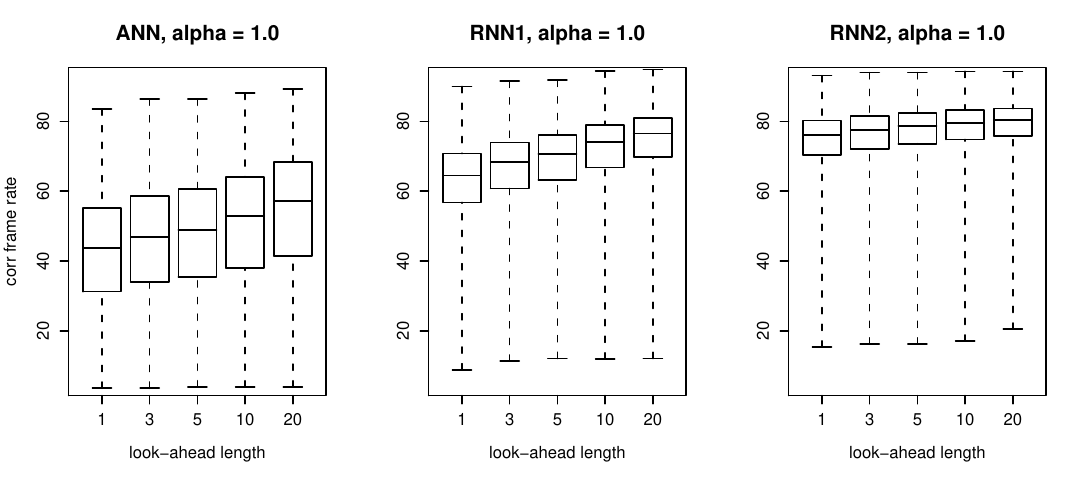}
\includegraphics[scale=0.8]{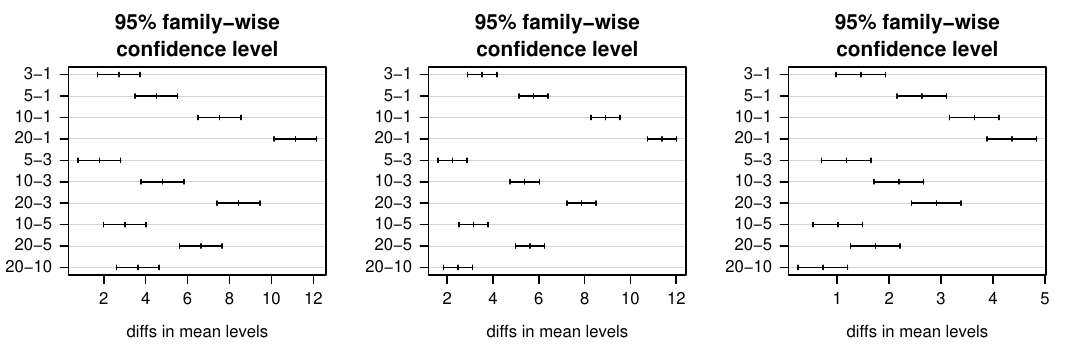}
\caption{Box-plots (top) and 95\% family-wise Tukey confidence intervals (bottom), alpha = 1.0} \label{Fig:alpha1}
\end{figure}

For $\alpha=1$, the LM specifies forced
alignment. The results in Figure~\ref{Fig:alpha1} indicate significant
increase of the correct frame classification rate with respect to the
look-ahead length, in all cases.
Finally, Table~\ref{Tab:alphasummary} shows the Tukey results in all
intermediate cases. These are not as regular as the ``wordlen''
results, revealing differences between the neural networks.
\begin{table}
\caption{Alpha test: Tukey HSD multiple comparison results} \label{Tab:alphasummary}
\begin{scriptsize}\begin{verbatim}
ANN
alpha  3-1   5-1   10-1  20-1  5-3   10-3  20-3  10-5  20-5 20-10
0.0     n     -     n     -     -     n     -     +     n     -
0.1     n     n     n     -     n     n     -     n     -     -
0.3     n     n     n     n     n     n     -     n     -     -
0.5     +     +     +     n     n     n     n     n     n     -
0.7     +     +     +     +     n     +     +     +     n     n
0.9     +     +     +     +     n     +     +     +     +     n
1.0     +     +     +     +     +     +     +     +     +     +
RNN1
alpha  3-1   5-1   10-1  20-1  5-3   10-3  20-3  10-5  20-5 20-10
0.0     +     +     +     +     n     +     +     +     +     n
0.1     +     +     +     +     n     +     +     n     n     n
0.3     +     +     +     +     +     +     +     +     +     n
0.5     +     +     +     +     +     +     +     +     +     n
0.7     +     +     +     +     +     +     +     +     +     n
0.9     +     +     +     +     +     +     +     +     +     +
1.0     +     +     +     +     +     +     +     +     +     +
RNN2
alpha  3-1   5-1   10-1  20-1  5-3   10-3  20-3  10-5  20-5 20-10
0.0     +     +     +     +     n     n     n     n     n     n
0.1     +     +     +     +     n     n     n     n     n     n
0.3     +     +     +     +     n     n     n     n     n     n
0.5     +     +     +     +     n     +     +     n     n     n
0.7     +     +     +     +     n     +     +     n     n     n
0.9     +     +     +     +     +     +     +     n     +     n
1.0     +     +     +     +     +     +     +     +     +     +
\end{verbatim}\end{scriptsize}
\end{table}

\subsection{Summary}
In both the ``wordlen'' and ``alpha'' tests, the statistic analysis
shows significant differences. A successive Tukey multiple comparison
test shows which couples of values of the look-ahead parameter
correspond to significant differences in the correct frame rate. The
results strongly depend on the MLP for short time dependences in the
recognition network (towards the phone loop condition). This
dependency fades when the condition is shifted towards the forced
alignment case.

\subsection{Confidence measure} \label{Sec:ConMeasResults}
\begin{figure}
  \centering
  \includegraphics[scale = 0.8]{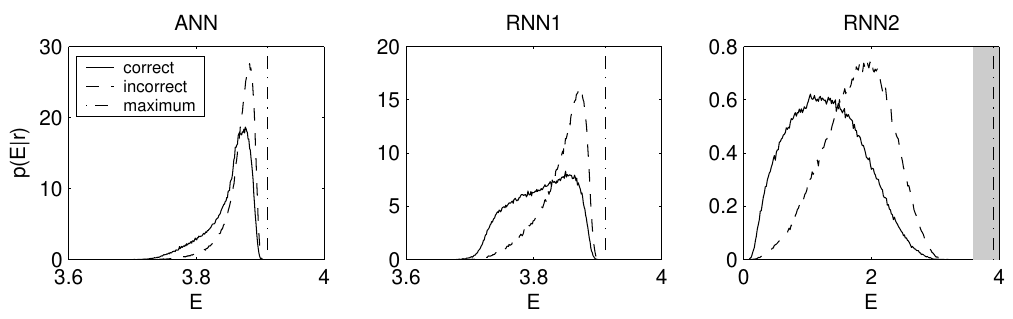}
  \caption{Entropy distribution for correct and incorrect
  classification}
  \label{Fig:EntropyMeasure}
\end{figure}

Figure~\ref{Fig:EntropyMeasure} shows the distribution of the entropy
for correct (continuous line) and incorrect (dashed line)
classification, and for the networks {\tt ANN} (left), {\tt RNN1} (centre), and
{\tt RNN2} (right). The vertical dashed-dotted lines indicate
the maximum entropy ($\log N$). In the rightmost plot, the shaded area
corresponds to the range chosen in the other two plots, and is used to
facilitate the comparison between the two conditions.
For the networks trained with $[0.1,0.9]$ targets ({\tt
  ANN} and {\tt RNN1}) the entropy is concentrated in the high range,
as explained in Section~\ref{Sec:confidence}. For {\tt RNN2} the
entropy range is larger (the network was trained with $[0,1]$
targets).

The prediction capabilities of the entropy as confidence
measure are however very similar for the recurrent networks. If we
consider a Maximum Likelihood decision, based on the conditional
entropy distributions, that leads to the minimum total
error in case of equal \emph{a priori} probabilities, we obtain the
results shown in Table~\ref{Tab:confidence}.
Note that the distributions shown in Figure~\ref{Fig:EntropyMeasure}
and the results in Table~\ref{Tab:confidence} are obtained on two
independent data sets created by splitting the test data into two
subsets of equal size.

\begin{table}
\caption{Prediction capabilities of the entropy as confidence measure (ML decision)} \label{Tab:confidence}
\begin{tabular}{l|cc|cc|c} \hline \hline
net & corr. accept & corr. reject & false accept & false reject &
tot. error \\ \hline
{\tt ANN} & 21.5\% & 37.2\% & 10.2\% & 31.1\% & 41.3\% \\
{\tt RNN1} & 32.1\% & 34.2\% & 17.7\% & 16.0\% & 33.8\% \\
{\tt RNN2} & 32.4\% & 33.9\% & 21.8\% & 11.9\% & 33.7\% \\ \hline \hline
\end{tabular}
\end{table}

\section{Discussion} \label{Sec:discussion}
The three factors considered in the experiments seem to strongly
interact in the decoding process. When the
language model (LM) is similar to a phone loop, most of the
information on the time evolution is provided by the multi-layer
perceptron. In this case differences emerge on the latency behaviour
of different neural network topologies. The static network ({\tt ANN})
produces irregular results when the look-ahead length $L$ is varied. The
dynamic models ({\tt RNN1} and {\tt RNN2}) show a slight improvement
with increasing $L$, that fades for higher values of $L$. The
look-ahead length for which no further improvement is achieved seems to be
lower for {\tt RNN2} than for {\tt RNN1}.

When the LM contains longer time dependencies, all acoustic models benefit
(to different extents) of longer look-ahead lengths. This can be
explained by noting that
\begin{itemize}
\item the Viterbi decoding makes use of time dependent information
  regardless of its source (transition model or dynamic neural
  network),
\item the information provided by the transition model and the dynamic
  neural network might overlap/conflict,
\item the look-ahead length needed to take full advantage of the
  Viterbi decoding is closely related to the length of the time
  correlations contained in the hybrid model (transition model or
  dynamic neural network).
\end{itemize}
Given these remarks, the results obtained here can be interpreted in
the following way.
The static model {\tt ANN}, providing no time dependent information,
takes advantage of the Viterbi decoding only for long time transition
models and long look-ahead. The more complex recurrent perceptron ({\tt
  RNN2}) provides information that partly overlaps with the transition model,
causing only limited improvements when the look-ahead is increased
(especially in the ``alpha'' test). The simpler recurrent perceptron
({\tt RNN1}) provides more limited time dependent information and
takes more advantage of the Viterbi decoding.

However, more specific tests should be
designed to support this interpretation, using, for example,
techniques from non-linear dynamics to analyse the dynamical behaviour
of the recurrent networks in details. Factors such as the target values during
training should also be excluded from the tests. The scoring method
used rises also questions on the kind of errors the system is affected
by in different conditions. It would be important, for example, to
investigate to which extent the errors are due to misclassification
of isolated frames or longer sequences, or to misalignment of
the segment boundaries.

\section{Conclusions} \label{Sec:conclusions}
The interaction of transition model, dynamic probability estimators
and look-ahead length in the decoding phase of a speech recognition
system has been analysed in this
paper. The results show how the dynamic information provided by the
recurrent multi-layer perceptrons does not always interact in a
constructive way with the transition model in Viterbi decoding.
With static MLPs, the use of longer look-ahead lengths is not beneficial when
the time dependencies in the language model are limited as in the
phone loop condition.
With recurrent MLPs, the benefit depends on the complexity of the network.

The frame-by-frame entropy proved to be a reasonably accurate confidence
measure. This measure is not strongly affected by the use of target
values in training other than [0,1].

\ack{This research was funded by the Synface European project
  IST-2001-33327 and carried out at the Centre for Speech Technology
  supported by Vinnova (The Swedish Agency for Innovation Systems),
  KTH and participating Swedish companies and organisations.}
\bibliographystyle{elsart-harv}
\bibliography{IEEEabrv,speechcomm2004}
\end{document}